\documentclass[twocolumn,showpacs,amsmath,amssymb,prl]{revtex4}
\usepackage{amsxtra}
\usepackage{amssymb}
\usepackage{amsmath}
\usepackage{graphicx}

\begin{document}
\title{Quantum illumination with Gaussian states}
\date{\today}
\author{Si-Hui Tan$^1$, Baris I. Erkmen$^2$\footnote{Now with the Jet Propulsion Laboratory, Pasadena, California 91109, USA}, Vittorio Giovannetti$^3$,\\ Saikat Guha$^2$\footnote{Now with BBN Technologies, Cambridge, Massachusetts, 02138, USA}, Seth Lloyd$^{2}$, Lorenzo Maccone$^4$, Stefano Pirandola$^2$, and\\ Jeffrey H. Shapiro$^2$}\email{jhs@mit.edu}
\affiliation{$^1$Massachusetts Institute of Technology, Department of Physics, Cambridge, Massachusetts 02139, USA\\ 
$^2$Massachusetts Institute of Technology, Research Laboratory of Electronics, Cambridge, Massachusetts 02139, USA \\ $^3$NEST-CNR-INFM \& Scuola Normale Superiore, Piazza dei
  Cavalieri 7, I-56126, Pisa, Italy \\ $^4$QUIT, Dipartimento di Fisica ``A. ~Volta'', Universita' degli studi di Pavia, via Bassi 6,
  I-27100 Pavia, Italy}
  
\begin{abstract} 
An optical transmitter irradiates a target region containing a bright thermal-noise bath in which a low-reflectivity object might be embedded.  The light received from this region is used to decide whether the object is present or absent.  The performance achieved using a coherent-state transmitter is compared with that of a quantum illumination transmitter, i.e., one that employs the signal beam obtained from  spontaneous parametric downconversion (SPDC).   By making the optimum joint measurement on the light received from the target region together with the retained SPDC idler beam, the quantum illumination system realizes a 6\,dB advantage in error probability exponent over the optimum reception coherent-state system.  This advantage accrues despite there being no entanglement between the light collected from the target region and the retained idler beam.  
  
\end{abstract}

\maketitle

Entanglement is a fundamental quantum mechanical resource, with potential applications to Heisenberg-limited precision measurements \cite{GLMscience}, teleportation \cite{teleport}, and quantum cryptography \cite{Ekert}.  Loss and noise, however, can quickly destroy this entanglement.  Recently, Lloyd \cite{QI}, building on the work of Sacchi \cite{Sacchi}, has shown that ``quantum illumination'' can reap substantial benefits, from the use of entanglement in target detection, despite the presence of entanglement-destroying loss and noise.  In Lloyd's quantum illumination paradigm, a photonic source creates $d$-mode maximally-entangled signal and ancilla beams each containing a single photon.  The signal beam irradiates a target region containing a very weak thermal-noise bath---with an average of $b \ll 1$ photons per mode---in which a low-reflectivity ($\eta \ll1$) object might be embedded.  The light received from this region---together with the retained ancilla beam---is then used to decide whether the object is present or absent.  Lloyd showed that quantum illumination, with the optimum joint measurement on the received light and the ancilla, achieves an effective signal-to-background ratio of $\eta d/b$, whereas optimum quantum reception of light received in response to transmission of a single unentangled photon has 
a much lower (for $d\gg 1$) signal-to-background ratio of $\eta/b$.  

The analysis in \cite{QI} was confined, primarily, to the vacuum plus single-photon manifold, wherein at most one photon arrives at the receiver during the measurement interval, regardless of whether the object of interest is absent or present in the target region.  Lloyd did briefly describe a microcanonical noise model whose quantum-illumination performance---for multi-photon signal and noise states with the same energy---again showed the factor of $d$ improvement in the signal-to-background ratio.  However the microcanonical model is rather restrictive, in that thermal-noise baths are in Gaussian states with Bose-Einstein distributed photon numbers.  In this Letter we will remedy that deficiency by providing a full Gaussian-state treatment of quantum-illumination target detection.  Our treatment will employ the exact quantum statistics for the entangled signal and idler beams obtained from continuous-wave (cw) spontaneous parametric downconversion (SPDC) in the absence of pump depletion \cite{LaserPhys}.  It will also use the standard model for the lossy, noisy bosonic channel \cite{channels}.  We will show that in a very lossy, very noisy environment, a low-brightness quantum illumination system enjoys a substantial improvement in effective signal-to-background ratio---which can translate into a very large reduction of the target-detection error probability---in  comparison to that achieved by a coherent-state transmitter of the same average photon number.  As seen in \cite{QI}, this performance advantage accrues to the quantum illumination system despite there being no entanglement between the light that is received from the target region and the retained idler beam.  Quantum illumination thus becomes the first example of an entanglement-based performance gain, in a full bosonic-channel setting, that survives entanglement-killing loss and noise.  

Consider an entangled signal and idler mode pair obtained from cw SPDC.  This mode pair---with annihilation operators $\hat{a}_S$ and $\hat{a}_I$---is in the entangled state with number-ket representation 
\begin{equation}
|\psi\rangle_{SI} = \sum_{n=0}^\infty \sqrt{\frac{\displaystyle N_S^n}{(N_S+1)^{n+1}}}\,|n\rangle_S|n\rangle_I,
\end{equation}
where $N_S$ is the average photon number per mode.  In the quadrature representation, $|\psi\rangle_{SI}$ is a zero-mean Gaussian state whose Wigner-distribution covariance matrix is
\begin{equation}
{\boldsymbol \Lambda}_{SI} = \frac{1}{4}\left[\begin{array}{cccc}
S  & 0 & C_q & 0 \\ 
0 & S  & 0 & -C_q \\
C_q & 0 & S & 0 \\ 
0 & -C_q & 0 & S
\end{array}\right], \label{quadent}
\end{equation}
where $S \equiv 2N_S + 1$ and $C_q \equiv 2\sqrt{N_S(N_S+1)}$.   
Viewed in this way, it is easily seen that $|\psi\rangle_{SI}$ has maximally-entangled quadratures, because the magnitudes of non-zero off-diagonal terms, $C_q$, in ${\boldsymbol \Lambda}_{SI}$ equal the maximum value allowed by quantum mechanics given the diagonal elements of that matrix.  Indeed, the upper limit on the magnitudes of these off-diagonal terms for a classical state, i.e., one with a proper $P$ representation so that the signal and idler modes are unentangled,  is $C_c\equiv 2N_S$.  Thus, for a low-brightness source, for which the average number of photons per mode is very low ($N_S \ll 1$), there is a very strong nonclassical signature in ${\boldsymbol \Lambda}_{SI}$ in that $C_q \gg C_c$ prevails.  

Suppose that we transmit the signal mode toward a spatial region that may or may not contain a weakly-reflecting target but, in either case, contains a bright thermal-noise bath.  Also suppose that the transmitter retains the idler mode, for a subsequent joint measurement to be made with the return from that target region.  Under hypothesis $H_0$ (target absent), the annihilation operator for the return from the target region will be $\hat{a}_R = \hat{a}_B$, where $\hat{a}_B$ is in a thermal state with average photon number $N_B\gg 1$.  Under hypothesis $H_1$ (target present), the return-mode's annihilation operator will be $\hat{a}_R = \sqrt{\kappa}\,\hat{a}_S + \sqrt{1-\kappa}\,\hat{a}_B$, where $\kappa \ll 1$ and $\hat{a}_B$ is now in a thermal state with average photon number $N_B/(1-\kappa)$.  Physically, this represents a very lossy ($\kappa \ll 1$) return from the target when it is present, combined with a very strong background contribution ($N_B \gg 1$) that, at the receiver, is independent of target absence or presence \cite{footnote1}.  

Under both hypotheses the joint state of the $\hat{a}_R$ and $\hat{a}_I$ modes is a zero-mean mixed Gaussian state, with conditional Wigner-distribution covariance matrices given by
\begin{equation}
{\boldsymbol \Lambda}_{RI}^{(0)} = \frac{1}{4}\left[\begin{array}{cccc}
B  & 0 & 0 & 0 \\ 
0 & B & 0 & 0\\
0& 0 &  S & 0 \\ 
0 & 0 & 0 & S
\end{array}\right]
\end{equation}
under $H_0$, and 
\begin{equation}
{\boldsymbol \Lambda}_{RI}^{(1)} = \frac{1}{4}\left[\begin{array}{cccc}
A  & 0 & \sqrt{\kappa}C_q & 0 \\ 
0 & A & 0 & -\sqrt{\kappa}C_q\\
\sqrt{\kappa}C_q & 0 & S & 0 \\ 
0 & -\sqrt{\kappa}C_q& 0 & S
\end{array}\right]
\end{equation}
under $H_1$, where $B\equiv 2N_B+1$ and $A \equiv 2\kappa N_S + B$.  
Neither of these conditional states is entangled.   This is clearly so when the target is absent, because the bath mode $\hat{a}_B$ is independent of the idler mode $\hat{a}_I$.  When the target is present the lack of entanglement when $N_B > \kappa$ follows from the magnitude of the off-diagonal elements, $\sqrt{\kappa}C_q$, in ${\boldsymbol \Lambda}^{(1)}_{RI}$ falling below the classical-state limit, $2\sqrt{N_S(\kappa N_S+N_B)}$, set by the diagonal elements in that matrix.  In other words, the conditional density operators, $\hat{\rho}^{(0)}_{RI}$ and $\hat{\rho}^{(1)}_{RI}$, for the return and idler modes under hypotheses $H_0$ and $H_1$ both have proper $P$ representations, and hence can be regarded as classically-random mixtures of coherent states.  

When the two hypotheses are equally likely, the minimum error probability decision rule for the quantum illumination receiver is as follows \cite{Helstrom}.  Measure $\hat{\rho}_{RI}^{(1)}-\hat{\rho}_{RI}^{(0)}$ and declare the target to be present if a non-negative outcome results, and otherwise declare the target to be absent.  Moreover, with $\{\gamma_n^{(+)}\}$ denoting the non-negative eigenvalues of
$\hat{\rho}_{RI}^{(1)}-\hat{\rho}_{RI}^{(0)}$, we have that the error probability of this optimum quantum receiver for the quantum-illumination transmitter is
\begin{equation}
\Pr(e) =  \left(1 - \sum_n \gamma_n^{(+)}\right)/2. \label{PeGeneral}
\end{equation}
Unfortunately, finding the eigenvalues of $\hat{\rho}_{RI}^{(1)}-\hat{\rho}_{RI}^{(0)}$ is a daunting task, so, as is often done in communication theory, we shall establish bounds on this error probability.   Before doing so, however, we pause to introduce two classical-state comparison cases for the quantum illumination system.  The first uses a coherent-state transmitter with average photon number $N_S$.  The second employs signal and idler modes that are in a joint classical state with only the signal being transmitted toward the target region, while the idler is retained for possible use in conjunction with return mode from the target region.  Optimum quantum reception will be assumed for both of these cases, so that they too employ difference of conditional density operator measurements and have error probabilities given by  \eqref{PeGeneral} in terms of the non-negative eigenvalues of their measurement operators.  

Based on \cite{QI}, which showed that quantum illumination is advantageous at low signal-to-background ratios, and the fact that the off-diagonal elements in ${\boldsymbol \Lambda}_{SI}$ asymptotically approach classical-state behavior for $N_S \gg 1$, we have chosen to study quantum illumination in the regime wherein $N_S \ll 1$ and $N_B \gg 1$.  Only a moment's thought is needed to conclude that the error probability in this operating regime will be very close to 1/2 for our quantum illumination system, as well as for the coherent-state and classical joint-state comparison cases.  So, to get to acceptably low error probabilities with the quantum illumination and coherent-state systems we will use $M \gg 1$ identical transmissions of the types described above, in conjunction with optimum joint quantum measurements.  Thus, the joint signal-idler state for the quantum illumination transmitter will be $\hat{\boldsymbol\rho}_{SI} = \otimes_{m=1}^M\hat{\rho}_{S_mI_m}$, where the modal signal-idler states are zero-mean and jointly Gaussian with the Wigner distribution covariance matrix from \eqref{quadent}, and the coherent-state transmitter will emit $\otimes_{m=1}^M|\sqrt{N_S}\rangle_m$, where $\{|\cdot\rangle_m\}$ are the modal coherent states.   The description of our general classical-state transmitter will be given later.

At this point, the quantum Chernoff bound \cite{Chernoff} comes to our rescue.  For optimum quantum discrimination between a pair of equally likely $M$-mode conditional density operators, $\hat{\boldsymbol\rho}^{(k)} = \otimes_{m=1}^M\hat{\rho}^{(k)}_m$ for $k=0,1$, with identical modal states under each hypothesis, viz., $\hat{\rho}^{(k)}_m = \hat{\rho}^{(k)}$ for $1 \le m \le M$ and $k= 0,1$, the quantum Chernoff bound places the following limit on the error probability:
\begin{equation}
\Pr(e) \le \frac{1}{2}e^{-M{\cal{E}}}
\equiv \frac{1}{2}\!\left[\min_{0\le s\le 1}\rm{tr}\!\left((\hat{\rho}^{(0)})^s(\hat{\rho}^{(1)})^{(1-s)}\right)\right]^M .
\end{equation}
This bound is exponentially tight, 
i.e., the error probability exponent, $-\ln[\Pr(e)]/M$, converges to ${\cal{E}}$ as $M\rightarrow \infty$.
Its potentially weaker $s=1/2$ version, known as the Bhattacharyya bound,
\begin{equation}
\Pr(e) \le \frac{1}{2}\!\left[{\rm tr}\!\left((\hat{\rho}^{(0)})^{1/2}(\hat{\rho}^{(1)})^{1/2}\right)\right]^M,
\end{equation}
will also be of interest in what follows, because it is more amenable to obtaining analytic results and because it is related to the lower bound on the error probability
\begin{equation}
\Pr(e) \ge \frac{1}{2}\!\left(1-\sqrt{1- \left[{\rm tr}\!\left((\hat{\rho}^{(0)})^{1/2}(\hat{\rho}^{(1)})^{1/2}\right)\right]^{2M}}\right),
\end{equation}
which, in general, is \em not\/\rm\  exponentially tight.  

Our quantum illumination and coherent-state transmitters both lead to modal density operators, under each hypothesis, that are Gaussian states.  Hence we can use the results of \cite{Pirandola} to evaluate Chernoff or Bhattacharyya bounds.  To do so we need three things:  the mean values of the relevant modes---$\hat{a}_R$ for the coherent-state transmitter, and $\hat{a}_R$ and $\hat{a}_I$ for the quantum illumination transmitter---under each hypothesis; the conditional Wigner-distribution covariance matrices for these relevant modes; and the symplectic diagonalizations of those conditional Wigner-distribution covariance matrices.  The symplectic diagonalization  of a $2K\times 2K$ dimensional covariance matrix $\boldsymbol \Lambda$ consists of a $2K\times 2K$ dimensional symplectic matrix $\boldsymbol S$ and a symplectic spectrum $\{\,\nu_k : 1\le k \le K\,\}$ that satisfy
\begin{eqnarray}
{\boldsymbol S}{\boldsymbol \Omega}{\boldsymbol S}^T &=& {\boldsymbol \Omega} 
\equiv \bigoplus_{k=1}^K \left(\begin{array}{cc} 0 & 1 \\[.1in]
-1 & 0 \end{array}\right) \\[.1in]
{\boldsymbol \Lambda} &=& {\boldsymbol S}\,{\rm diag}(\nu_1,\nu_1,\nu_2,\nu_2,\ldots,\nu_K,\nu_K)
{\boldsymbol S}^T,
\end{eqnarray} 
where diag($\cdot,\cdot,\ldots,\cdot)$ denotes a diagonal matrix with the given diagonal elements.  

For the coherent-state transmitter we have that ${\rm tr}(\hat{\rho}^{(0)}\hat{a}_R) = 0$, 
${\rm tr}(\hat{\rho}^{(1)}\hat{a}_R) = \sqrt{\kappa N_S}$, ${\boldsymbol \Lambda}^{(0)} = 
{\boldsymbol \Lambda}^{(1)} =  {\rm diag}(B/4,B/4)$.  It follows that ${\boldsymbol S}$ is the 2-D identity matrix and $\nu_1 = B/4$, so that the results in \cite{Pirandola} lead to the quantum Chernoff bound (which turns out to be the Bhattacharyya bound)
\begin{eqnarray}
\Pr(e)_{\rm CS} &\le& e^{-M\kappa N_S\left(\sqrt{N_B+1}-\sqrt{N_B}\right)^2}/2 
\label{CSchernoff}\\[.1in]
& \approx&  e^{-M\kappa N_S/4N_B}/2,\mbox{ when $N_B \gg 1$.}
\end{eqnarray}
Because \eqref{CSchernoff} is also the Bhattacharyya bound, we have 
\begin{eqnarray}
\lefteqn{\Pr(e)_{\rm CS} \ge } \nonumber \\[.1in]
&&  \frac{1}{2}\!\left(1-\sqrt{1- e^{-2M\kappa N_S(\sqrt{N_B+1} - \sqrt{N_B})^2}}\right)
\label{CSlower1} \\[.1in]
&\approx & e^{-M\kappa N_S/2N_B}/4, 
\label{CSlower2}
\end{eqnarray}
when $N_B \gg 1$ and $M\kappa N_S/2N_B \gg 1$.

For the quantum-illumination transmitter, the 4-D identity matrix is the symplectic matrix needed for the diagonalization of ${\boldsymbol \Lambda}^{(0)}_{RI}$, and
\begin{equation}
{\boldsymbol S} = \left(\begin{array}{cc}
{\bf X}_+ & {\bf X}_- \\[.1in] {\bf X}_- & {\bf X}_+\end{array}\right),
\end{equation}
is the symplectic matrix needed to diagonalize ${\boldsymbol \Lambda}^{(1)}_{RI}$.  Here, 
${\bf X_\pm} \equiv {\rm diag}(x_\pm,\pm x_\pm)$ with 
\begin{equation}
x_\pm \equiv \sqrt{\frac{A+S \pm\sqrt{(A+S)^2 - 4\kappa C_q^2}}
{2\sqrt{(A+S)^2 - 4\kappa C_q^2}}}.
\end{equation}
 The associated symplectic spectra are $\nu_1 = B/4, \nu_2 = S/4$ under $H_0$, and 
\begin{equation}
\nu_k = \left[(-1)^k(S - A) + \sqrt{(A+S)^2 - 4\kappa C_q^2}\right]/8,
\end{equation}
for $k = 1,2$, under $H_1$.  These diagonalizations can be employed to derive an analytic expression for the Bhattacharyya bound on the error probability achieved with quantum illumination.  Unfortunately, that expression is far too long to exhibit here.  In Fig.~1 we compare the coherent-state system's Chernoff bound from \eqref{CSchernoff} and its lower bound from \eqref{CSlower1} with the quantum illumination system's Bhattacharyya bound when $\kappa = 0.01, N_S = 0.01$, and $N_B  = 20$.  We see that the quantum illumination system's error probability 
\em upper\/\rm\ bound---at a given $M$ value---can be orders of magnitude lower than the error probability \em lower\/\rm\ bound for  coherent-state light.  
\begin{figure}[h]
\begin{center}
\includegraphics[width=2.25in]{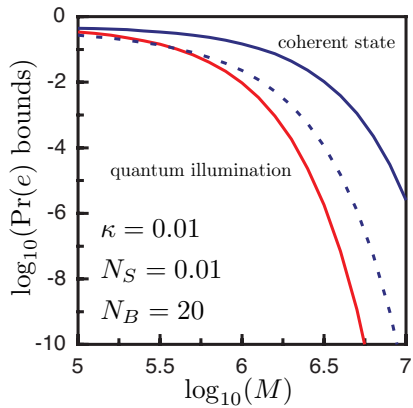}
\end{center}
\vspace*{-.2in}
\caption{Upper bounds (solid curves) on the target-detection error probabilities for coherent-state (Chernoff bound) and quantum illumination (Bhattacharyya bound) transmitters with $M$ transmitted modes each with $N_S = 0.01$ photons on average when $\kappa = 0.01$ and $N_B = 20$.  Also shown is the lower bound (dashed curve) for the coherent-state case, which, see below, also applies to \em all\/\rm\ classical-state transmitters with $\sum_{m=1}^M \langle \hat{a}_{S_m}^\dagger \hat{a}_{S_m}\rangle = MN_S$.}
\end{figure}

To show that the advantage afforded by quantum illumination extends well beyond the specific example chosen for Fig.~1, we have used an algebraic computation program to obtain the following approximate form for the quantum illumination transmitter's Bhattacharyya bound when $\kappa \ll 1, N_s \ll 1$ and $N_B \gg 1$:
\begin{equation}
\Pr(e)_{\rm QI} \le e^{-M\kappa N_S/N_B}/2.
\label{PeQI}
\end{equation}
Comparing this low-brightness quantum illumination result to the Chernoff bound for the coherent-state transmitter in the same lossy, noisy environment reveals that the use of entanglement at the transmitter improves the error probability exponent by a factor of 4, i.e., by 6\,dB.  One might wonder whether there is some other classical-state source that can match the performance of the quantum illumination transmitter in this regime.  To show that such is not the case, we now develop the perfect-measurement bound on the performance of \em any\/\rm\ classical-state transmitter.  Consider the general classical-state (signal and idler) transmitter whose output state is 
\begin{equation}
\hat{{\boldsymbol \rho}}_{SI} \equiv \int\!\int\!{\bf d}^2{\boldsymbol \alpha}_S{\bf d}^2{\boldsymbol \alpha}_I\,P({\boldsymbol \alpha}_S,{\boldsymbol \alpha}_I)
|{\boldsymbol \alpha}_S\rangle_S|{\boldsymbol \alpha}_I\rangle_I{}_S\langle{\boldsymbol \alpha}_S|
{}_I\langle{\boldsymbol \alpha}_I|,
\end{equation}
where $\{|{\boldsymbol \alpha}_j\rangle_j\}$ for $j = S,I$ are the $M$-mode coherent states of the signal and idler and $P({\boldsymbol \alpha}_S,{\boldsymbol \alpha}_I)$ is a probability density function over $\{\,{\boldsymbol \alpha}_j: j = S,I\,\}$ satisfying $\sum_{m=1}^M \langle \hat{a}_{S_m}^\dagger \hat{a}_{S_m}\rangle = MN_S$.  Minimum error probability reception when the signal beam irradiates the target region and a joint measurement is made of the return from the target region and the retained idler is bounded from below by the error probability achieved, using this transmitter, when ${\boldsymbol \alpha}_S$ is known to the receiver.  A noisy \em estimate\/\rm\ of ${\boldsymbol \alpha}_S$ can be obtained from heterodyne detection of ${\boldsymbol \alpha}_I$, hence our lower bound is correctly termed a ``perfect-measurement'' bound.  More importantly, given knowledge of ${\boldsymbol \alpha}_S$, the classical-state system reverts to being a coherent-state transmitter, so that \eqref{CSlower1}, convexity, and the average photon-number constraint imply that $\Pr(e)_{\rm class}$ obeys the same lower bound as $\Pr(e)_{\rm CS}$, 
showing that when $N_B \gg 1$  \em all\/\rm\ classical-state transmitters have error probability exponents that are at least 3\,dB inferior to that of quantum illumination with the same  $\sum_{m=1}^M \langle \hat{a}_{S_m}^\dagger \hat{a}_{S_m}\rangle$ value.

Given that $N_B =\kappa$ suffices to destroy the entanglement between the $\hat{a}_{S_m}$ and $\hat{a}_{R_m}$ modes under hypothesis $H_1$, it behooves us to find a physical explanation for the performance gain provided by quantum illumination.  In our view the explanation is as follows.  When $N_S \ll 1$, the entangled $\hat{a}_{S_m}$ and $\hat{a}_{I_m}$ modes have a cross correlation that greatly exceeds what is permitted for a classical state.  Although the corresponding cross correlation for $\hat{a}_{R_m}$ and $\hat{a}_{I_m}$ when the target is present \em is\/\rm\ within classical bounds, there is no classical \em input\/\rm\ state for $\hat{a}_{S_m}$ and $\hat{a}_{I_m}$ of the same average photon number that can produce a close approximation to this output state.  Of course, a single-mode transmission with $N_S \ll 1$, $\kappa \ll 1$, and $N_B \gg 1$ cannot yield a low error-probability decision.  However, the \em joint\/\rm\ measurement over $M \gg 1$ such modes can.  Moreover, no special joint-state performance enhancement is realized with the coherent-state transmitter, because the product state $\otimes_{m=1}^M |\sqrt{N_S}\rangle_m$ is merely a coherent state $|\sqrt{MN_S}\rangle$ of a different mode.  In summary, we must be careful \em not\/\rm\ to dismiss the value of using entangled resources just because the application scenario is lossy and noisy.

This research was supported by W. M. Keck Foundation Center for Extreme Quantum Information Theory and by the DARPA Quantum Sensors Program.


\begin{thebibliography}{2}
\bibitem{footnote1}For $\kappa \ll 1$ there is very little difference in the $\langle \hat{a}_B^\dagger\hat{a}_B\rangle$ values under $H_0$ and $H_1$. These values ensure that a non-vacuum transmitter must be used to detect target presence.
\end{thebibliography}
\end{document}